\documentclass[conference]{IEEEtran}
\IEEEoverridecommandlockouts

\usepackage{titlesec}
\usepackage{comment}

\titlespacing{\section}{0pt}{0.5ex plus 0ex minus 0.5ex}{0.5ex plus 0ex minus 0.5ex}

\titlespacing{\subsection}{0pt}{0.5ex plus 0ex minus 0.5ex}{0.5ex plus 0ex minus 0.5ex}

\titlespacing{\subsubsection}{0pt}{0.5ex plus 0ex minus 0.5ex}{0.5ex plus 0ex minus 0.5ex}

\usepackage{cite}
\usepackage{amsmath,amssymb,amsfonts}
\usepackage{algorithmic}
\usepackage{graphicx}
\usepackage{textcomp}
\usepackage[table, dvipsnames]{xcolor}
\usepackage{enumitem}
\usepackage{listings}
\usepackage{tikz}
\usepackage{tcolorbox}
\usetikzlibrary{positioning}
\usepackage{balance}
\usepackage{url}

\usepackage{array}
\usepackage{booktabs}  % For \toprule, \midrule, \bottomrule
\usepackage{array}     % For column specifications
\usepackage{threeparttable}  % For table notes (if you want integrated footnotes)
\usepackage{multirow}

\usepackage{stfloats} 
\usepackage{ragged2e}   % For \RaggedRight

\usepackage[ruled,vlined,linesnumbered]{algorithm2e}
\usepackage{soul}
\usepackage{caption}
\SetAlFnt{\small}

\usepackage{xcolor}
\usepackage{colortbl}
\definecolor{lightgray}{gray}{0.9}
\definecolor{headercolor}{RGB}{230,230,250}

\def\BibTeX{{\rm B\kern-.05em{\sc i\kern-.025em b}\kern-.08em
    T\kern-.1667em\lower.7ex\hbox{E}\kern-.125emX}}

\begin{document}\setlength\emergencystretch{1.5em}
\title{Feature-Centric Methodology for Analyzing Cross-Chain NFT Migration Compatibility}

\author{\IEEEauthorblockN{Mohd Sameen Chishti}
\IEEEauthorblockA{\textit{Department of Computer Science} \\
\textit{NTNU}\\
Trondheim, Norway \\
mohd.s.chishti@ntnu.no}
\and
\IEEEauthorblockN{Damilare Peter Oyinloye}
\IEEEauthorblockA{\textit{Department of Computer Science} \\
\textit{NTNU}\\
Trondheim, Norway \\
peter.d.oyinloye@ntnu.no}
\and
\IEEEauthorblockN{Jingyue Li}
\IEEEauthorblockA{\textit{Department of Computer Science} \\
\textit{NTNU}\\
Trondheim, Norway \\
jingyue.li@ntnu.no}
}

\maketitle

\begin{abstract}

As blockchain ecosystems fragment across multiple chains optimized for cost, security, and other parameters, creators, owners, and marketplaces increasingly face the need to migrate non-fungible tokens (NFTs) to alternative blockchain platforms without compromising their functionality or value. Cross-chain NFT migration refers to the process of transferring digital assets along with their associated functionalities and guarantees between distinct blockchain platforms. However, architectural divergences among these platforms introduce critical challenges, often resulting in features that fail to behave as intended. While protocol-level mechanisms can coordinate data transfer, they are insufficient to resolve deeper compatibility issues arising from fundamental differences in state organization, transaction execution, and ownership representation. Thus, the critical challenge lies in predicting which NFT features can be preserved, which require redesign, and which are fundamentally incompatible, prior to undertaking costly migration attempts. To address this challenge, we first derive a tailored four-layer NFT architecture based on standard blockchain stacks, distinguishing cryptographic, state-management, transaction-processing, and ownership primitives, with explicit upward dependencies. Building on this architecture, we conceptualize an NFT as a bundle of features and define successful cross-chain NFT migration as the preservation of these features.
Grounded in this model, we propose a four-phase migration analysis methodology comprising source feature specification, primitive-level dependency mapping, target platform profiling, and compatibility assessment, which classifies each feature as natively preserved, partially mismatched, or completely mismatched.  We evaluate this methodology through a proof-of-concept analysis of Ethereum-to-Solana NFT migration, identifying several incompatibility issues that hinder seamless NFT migration.
% The methodology enables evidence based migration planning by revealing which features preserve natively, partial mismatch, or complete mismatch.

\end{abstract}


\section{Introduction}

The blockchain ecosystem has evolved into a highly fragmented landscape of multiple platforms, each optimized for different trade-offs in terms of throughput, cost, decentralization, and functionality \cite{MoralisBlockchains}. This has led to a situation where blockchains are isolated with incompatible architectures, consensus mechanisms, and data standards, hindering seamless exchange of assets and information \cite{LI2025100286}. Thus, cross-chain transfers have become a critical research and development area to address these issues \cite{10.1145/3727648.3727685}. This is also evidenced by the existence of over a hundred cross-chain bridge protocols and over \$$41$Bn in cross-chain transactions in $2024$  \cite{InterchainReport2024}.

Non-fungible tokens (NFTs) represent one of the most challenging asset categories for cross-chain mobility due to their complex implementation requirements \cite{GUO2024101968,ALI2023122248}. In comparison to fungible tokens, where cross-chain migration mechanisms need to track only numeric balances, NFT migration depends on a specific pattern for identity assignment, ownership representation,  provenance, and metadata management. A cross-chain NFT migration requires transferring not only ownership but also all the properties and behaviors that define the NFT, including its identity, ownership history, metadata references, and associated rules.

Recent attempts at NFT migration underscore the practical complexity of the process. Even when tokens are successfully transferred using burn-and-mint or lock-and-wrap mechanisms \cite{10.1145/3678890.3678894}, post-migration analyses reveal functional failures. For instance, inconsistencies in CloneX metadata, trait degradation in Pixelmon, and trait randomization in WhaleNFT, along with a 3 million USD grant to support the y00ts migration, highlight technical constraints that economic incentives alone could not overcome \cite{BeInCrypto, Tekedia, Hypebeast, WhaleNFT}. The recurring pattern in these cases is that ownership migrates successfully, but not all features are preserved. Bridges can ensure that only one active representation of a token exists across chains. However, the behavioral aspects of NFTs, such as identity, ownership representation, metadata linkage and interpretation, and trait behavior, can change on the target blockchain.
For example, a collection that uses sequential numeric token IDs (\#1 signifying the founder's token which often entails premium pricing) cannot preserve this meaning on a platform where token identity must be cryptographically derived from seeds. 
Similarly, a collection that relies on a centralized ownership registry cannot be reproduced on a platform that mandates distributed token accounts.
We refer to these situations as NFT migration mismatches. In this work, our focus is  on permissionless public blockchain and extension to permissioned ledgers is left for future work.

Existing bridge protocols largely treat NFTs as opaque payloads and do not explicitly verify whether the destination platform can support the identity, ownership, metadata, and other feature-level behaviors on which the NFTs depend at the source. The ability to preserve these features is instead determined by deeper architectural choices within each blockchain-choices that current bridge protocols and standards do not explicitly model. Such challenges motivate us to answer the research question:
\emph{How can we systematically analyze cross-chain NFT migration to identify potential migration mismatches?}

To address the research question, we first derive a layered NFT architecture from the standard blockchain stack, identifying the core components necessary to implement NFT features. 
This provides a structured framework for analyzing the platform capabilities on which NFT features depend. Second, we analyze the inter-layer dependencies that shape NFT implementations, identifying how lower-layer design choices, such as addressing schemes, state organization, and transaction execution models,  constrain feature preservation during migration. Next, we introduce a systematic four-phase migration analysis methodology that maps source NFT features to their architectural requirements and assesses whether these requirements can be met on the target platform. The methodology utilizes the layered NFT architecture and the inter-layer dependencies of features, which enable a pre-migration assessment of compatibility. This allows developers to determine which features will be preserved after migration, which require redesign, and which face architectural impossibility. The methodology is evaluated through a proof-of-concept case study analyzing Ethereum-to-Solana NFT migration. The analysis identifies three potential partial mismatches and one complete mismatch if NFTs are migrated from Ethereum to Solana.

The remainder of this paper is organized as follows: Section \ref{sec:background} reviews NFTs and their core requirements and features. Section \ref{sec:cross-chain-nft-migration} outlines cross-chain migration challenges. Section \ref{sec:related-work} summarizes related work. Section \ref{sec:NFTArchitecture} introduces our four-layer NFT architecture and examines inter-layer dependencies. Section \ref{sec:migration-analysis} details our four-phase migration analysis methodology. Section \ref{sec:case-study} presents the evaluation design and results. Section \ref{sec:discussion} discusses the results and \ref{sec:conclusion} concludes the paper.

\section{NFT Requirements and Features}\label{sec:background}

%\subsection{NFT}\label{sec:DefiningNFT}

An NFT is a blockchain-based software abstraction that represents a unique, non-interchangeable asset and provides verifiable ownership and transferability \cite{espel2025cross}. Unlike the fungible token, where only numeric balances matter, an NFT carries a unique identity, ownership records, metadata links, and provenance history. It may also have some rules for economic considerations, like royalties for the creator. 

%A migration that may preserve identity but break metadata linkage or  transfer ownership correctly but looses the royalty enforcement can not be said as truly succeeded. 
To effectively analyze NFT migration issues, we must first establish a clear definition of what constitutes an NFT and identify the features that govern its behavior. Based on widely accepted definitions (i.e., NFTs as unique, non-fungible digital identifiers recorded on a blockchain, designed to certify ownership and authenticity while enabling the transfer of that ownership), and established NFT standards such as ERC-721 \cite{ERC721}, Solana's SPL token and Metaplex Metadata standards \cite{SolanaTokens,metaplex-token-metadata}, and similar specifications on other platforms \cite{FlowCadence, TezosMichelson}, we distill four core requirements necessary for successful migration across blockchains: (i) uniqueness of token identity \cite{ERC721,KAISTO2024105996}, (ii) exclusive ownership \cite{KAISTO2024105996,hofstetter2022crypto,10363651}, (iii) provenance-preserving transferability \cite{ALAVI2025610}, and (iv) association with descriptive metadata \cite{10.1007/978-3-031-87775-9_13}.   
\begin{itemize}[leftmargin=*]
 
    \item \textbf{Uniqueness:} Each token instance has a globally unique on-chain identifier that distinguishes it from all other tokens on the platform. 
    \item \textbf{Exclusive ownership:} At any point in time, exactly one blockchain account controls the token, as recorded in the ledger state.
    \item \textbf{Transferability with provenance:} The abstraction provides operations that transfer ownership between accounts and append to an immutable transaction history that records the sequence of owners.
    \item \textbf{Metadata association:} The token maintains an explicit association to descriptive metadata that links it to the digital or real-world asset or rights it represents, either through on-chain storage or references to off-chain storage.
\end{itemize}

These requirements manifest as concrete features in NFT implementations. Table \ref{tab:nftFeatures} shows the relationship between requirements and features. These are the core features for an NFT. We denote this set by $F_{\mathit{core}}$ = \{\text{identity mechanism}, \text{ownership representation}, \text{transfer logic}, \text{metadata linkage}\}. However, the NFT may have extended features in addition to the $F_{\mathit{core}}$ such as royalty enforcement, batch operation, soul-bound capabilities, and many more. We denote the set of such extended features by $F_{\mathit{ext}}$. The content of $F_{\mathit{ext}}$ depends on the application. Some platforms also support shared or fractional ownership. Such models can be included in $F_{\mathit{ext}}$ and analyzed using the same methodology. Throughout this paper, we treat an NFT implementation as a bundle of features $F = F_{\mathit{core}} \cup F_{\mathit{ext}}$. Our migration analysis begins with four core features that establish the minimal baseline, while incorporating additional features when they are relevant to the specific use case. A migration can be considered successful only if all features preserve their observable behavior on the target platform. Migration mismatches occur when one or more features fail to reproduce their source-chain behavior. %We now present a discussion on cross-chain bridges and how the NFT migration are performed.

\begin{table}
\centering
\caption{NFT Features}
\label{tab:nftFeatures}
\resizebox{0.4\textwidth}{!}{
\small
\begin{tabular}{p{2.5cm}p{1.5cm}p{3cm}}
\toprule
\textbf{Feature} & \textbf{Requirement } & \textbf{Description} \\
\midrule
Identity mechanism & Uniqueness &How tokens are assigned identifiers \\
Ownership representation & Exclusive Ownership & How current owner is recorded \\
Transfer logic &Transferability, Provenance & How ownership changes and history is logged \\
Metadata linkage & Metadata Association & How descriptive data is stored or referenced \\
\bottomrule

\end{tabular}
}
\end{table}

\section{Cross-Chain NFT Migration Challenges}\label{sec:cross-chain-nft-migration}

Cross-chain NFT migration refers to the process of re-establishing the identity, ownership, metadata, and behavioral guarantees of an NFT on a different blockchain platform while maintaining semantic continuity with the original token. The migration process needs to preserve the core properties of uniqueness, exclusive ownership, provenance record, metadata association, and other behavioral features that define the NFT itself. The migration process is challenging because the NFT is not a static data object, but rather is governed by the concrete implementations in different layers of the NFT architecture and their underlying choices of primitives. Thus, moving the NFT from one chain to another requires determining whether the target platform can replicate the semantic guarantees provided by the source platform. For example, if the source platform uses a sequential numeric token identifier enabled by a global tree primitive providing a centralized data structure to store the address mapping, and the target platform requires a deterministic cryptographically determined address for parallel executions, the numeric identity semantic can not be transferred without auxiliary mapping infrastructure because the destination does not support an arbitrary identity feature. 

Recent examples demonstrate that this architectural compatibility issue has significant economic and technical implications. The y00ts collection, for example, accepted a US\$3\, million non-equity grant to migrate it from Solana to Polygon, but after experiencing weak liquidity and market activity on the destination ecosystem, the project announced a second migration from Polygon to Ethereum and returned the full grant to Polygon Labs \cite{BeInCrypto,y00tsEthReturn}. Although the bridge functioned as intended, the migrated NFTs did not enjoy equivalent economic functionality because the surrounding infrastructure and user base differ across platforms.

In another incident, RTFKT’s CloneX collection temporarily lost visible artwork when its Cloudflare-hosted media endpoint was disabled. While the tokens remained on-chain, their images disappeared until the project migrated metadata to the decentralized storage network Arweave \cite{Tekedia}. This case illustrates how NFT behavior also depends on storage and networking primitives beyond the core ledger. Cross-chain designs can even change NFT semantics by design. Whale.io's bridge from TON to Solana burns original NFTs and mints new ones on Solana, assigning traits at random so that individual identity and rarity may change even though project branding remains the same \cite{WhaleNFT}. The randomization of traits can destroy the rarity distribution of token which in turn determine the token price.  In all of these examples, token balances move successfully, yet important NFT features are lost, altered, or weakened.

\section{Related Work}\label{sec:related-work}

Most of the work on cross-chain systems has so far focused on efficient cross-chain token transfers \cite{GUO2024101968}, security and vulnerability patterns in cross-chain bridges \cite{10.1145/3696429}, and risks introduced when migrating smart contracts between platforms \cite{blockchains2040018,zheng2023depth}. 
%In summary, the central challenge in cross-chain NFT migration is not whether tokens can be transported between chains, but whether the
%destination architecture can reproduce the feature-level behavior of the source implementation. Existing bridge mechanisms and NFT
%interoperability standards ensure transport, uniqueness and message verification, but treat NFTs largely as opaque payloads and do not
%analyze feature compatibility. We refer to any feature whose behavior cannot be reproduced on the target platform as a migration
%mismatch. In the next section, we build on the NFT definition and feature list from Section \ref{sec:DefiningNFT} and the architectural model from Section \ref{sec:NFTArchitecture} to formalize migration as a compatibility problem between feature requirements and platform capabilities, and to present a systematic methodology for analyzing migration mismatches.
Cross-chain bridges have emerged as the primary mechanism for transferring assets between blockchains, typically through \emph{burn-and-mint} or \emph{lock-and-wrap} schemes \cite{10.1145/3696429}. In \emph{burn-and-mint} schemes, the token on the source chain is
irreversibly destroyed, and a cryptographic proof of this burn is submitted to the destination chain, which then mints a new
representation of the asset \cite{9657007,sober2023decentralized}. In \emph{lock-and-wrap} schemes, the source-chain token is locked in an escrow or custodial contract and a wrapped representation is issued on the destination chain, backed by the locked original token \cite{9793325}.
Both schemes enforce a one-to-one relationship between the source and destination representations, relying on the bridge infrastructure to ensure that proofs are valid and not replayed.

Many existing cross-chain bridges primarily target fungible tokens and  focus on securely transferring balances and messages across chains, with NFTs treated largely as payloads attached to these mechanisms.  Some studies have started discussing NFT-specific interoperability mechanisms. For example, the ICS-721 standard in the Cosmos ecosystem defines an IBC-compatible format for inter-chain NFT transfers, including packet formats and metadata fields, and is compatible with ERC-721-style representations \cite{10202634,kwon2018network}. The Wormhole’s NFT bridge between Ethereum and Solana \cite{11185063}, and LayerZero’s ONFT standards for omnichain NFTs \cite{lawrence2025bridging} define protocols for locking or burning source-chain tokens and minting corresponding representations on destination chains. The IETF Secure Asset Transfer Protocol (SATP) defines gateway-based transfer mechanisms for both fungible and non-fungible tokens but it places compatibility analysis outside its scope \cite{ietf_satp_about}. Similarly, interoperability efforts within the Linux Foundation Decentralized Trust, such as Hyperledger Cacti focus on transfer infrastructure rather than feature-level compatibility \cite{cacti_github}. Across all of these efforts, the NFT is still treated as an opaque payload attached to balance and message transfer across chains.

% Although these mechanisms address transport, uniqueness, and message verification across heterogeneous chains, they also treat the NFT as an opaque payload attached to balance and message transfer across chains. 

In summary, existing studies are limited in that they do not clearly identify which NFT features, such as identity semantics, ownership structures, metadata behavior, and royalty mechanisms, can be accurately replicated on the destination platform.

\section{Architectural Model of NFT}\label{sec:NFTArchitecture}

To analyze feature-level migration mismatches, we must develop a precise understanding of how NFT features are implemented on each chain. Accordingly, we start with proposing a customized NFT architectural model grounded in well-established blockchain frameworks, complemented by a detailed analysis of the inter-layer dependencies within this architecture. This model enables us to trace how various features depend on underlying platform choices, thereby facilitating the identification and analysis of migration mismatch problems.

\subsection{NFT Architectural Model}\label{sec:nft:layerarchitecture}

%We have already established the feature-based view of NFTs in Section \ref{sec:DefiningNFT}. Now a question arises that what enables these features to exist on a particular blockchain platform? An NFT feature like sequential numeric id or cryptographically defined id  does not appear out of nothing, instead, it is built on the capabilities provided by the underlying blockchain standard. A platform that lacks or have different design choice at architecture level can not implement certain features in comparison to some other platform regardless how the code for NFT is written. 

%To analyze which features are possible on which platforms, we adopt a layered view of blockchain architecture tailored for NFT analysis.  
Blockchain researchers argue that standard blockchain stack can be divided into network and consensus, data, execution and application layers \cite{10.1007/978-3-319-55699-4_34,8487348, 9402747,TABATABAEI2023100575,9239372}. However, we observed that NFT features characterized in Table \ref{tab:nftFeatures} do not directly depend on the network or consensus rules. The NFT directly avail itself the generic services such as authenticated transactions, deterministic ordering and finality. So, we treat the network and consensus as implicit foundation and focus on the layers that characterize the NFT feature behavior. Next, we partition the remaining concerns into four functional layers by grouping design choices according to which aspect of NFT behavior they can affect. The cryptographic layer contains choices about keys, signatures and identifiers that shape how NFT identities and user accounts are defined. The state-management layer contains choices about how NFT
state and metadata are stored and enumerated. The transaction-processing layer contains choices about how NFT-related transactions are processed.
The ownership and capability layer contains choices about how control, transfer rules and additional features (such as royalties) are expressed.

\begin{figure}
    \centering
    \includegraphics[scale=0.5]{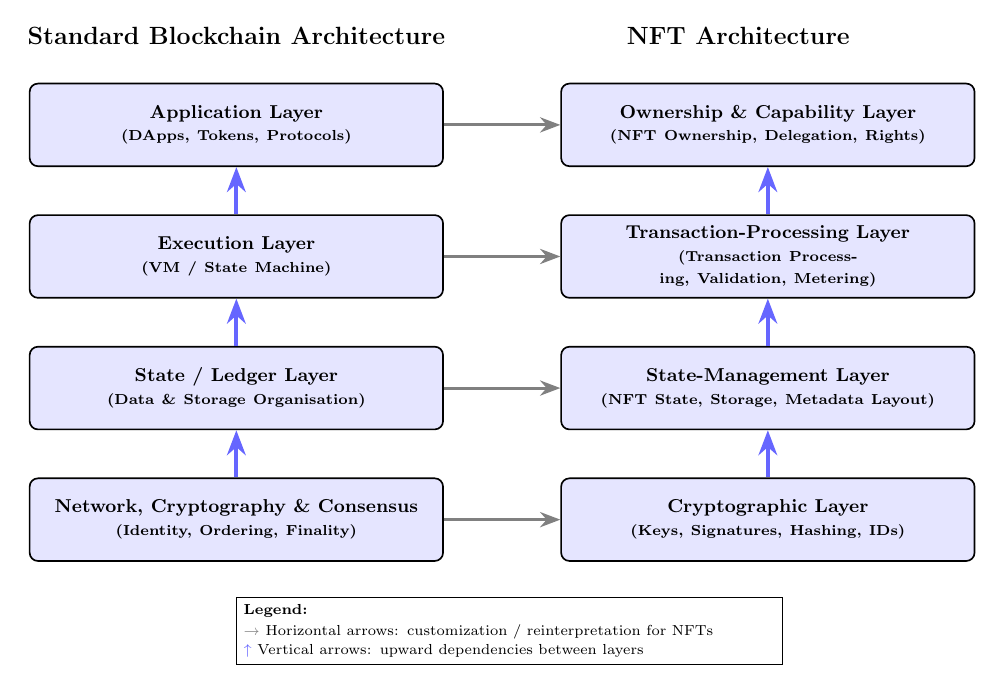}
    \caption{Standard Blockchain Architecture adapted from layered models \cite{10.1007/978-3-319-55699-4_34,9239372} and derived NFT Architecture}
    \label{fig:layeredArchitecture}
\end{figure}

We validated this decomposition against major NFT-supporting platforms, Ethereum (ERC-721, ERC-1155), Solana (SPL Token with Metaplex metadata), Flow (Cadence NonFungibleTokens, MetadataViews), and Tezos (FA2 contracts in Michelson) \cite{ERC721, 10.1007/978-3-031-82427-2_32, SolanaTokens,metaplex-token-metadata,FlowCadence, TezosMichelson}. These platforms together cover the dominant NFT ecosystems by market share and span distinct architectural styles (account-based EVM, parallel account-based runtime, resource-oriented smart contracts, and Michelson/FA2). In each case, observable differences in identity, ownership representation, and metadata handling can be traced to differences in one or more of these four layers. Fig. \ref{fig:layeredArchitecture} illustrates the four-layer NFT architecture and shows how it is derived from the standard blockchain
stack.
%We therefore use this model as our NFT architecture in the remainder of the paper.  We now describe each layer in detail.

\begin{itemize}[leftmargin=*]
    \item \textbf{Cryptographic Layer:} It provides the mechanisms for identity, integrity, and verification. The primitives  in the layers include signature scheme, hashing scheme, and address-derivation rule. These choices determine which public keys and signatures are valid, how content-addressed metadata and proofs are constructed, and whether token identifiers can be arbitrarily assigned or must be derived deterministically from seeds or scripts.

    \item \textbf{State-management layer:} The state-management layer specifies how the global blockchain state is
    structured, persisted, and priced. Its design dimensions include the organization of state, the semantics about data persistence, pricing model of stored data, and query capabilities for lookup. These design choices determine how the state  including identity mappings, ownership records, metadata references can be encoded, retained and retrieved.

    \item \textbf{Transaction-processing layer:} The transaction-processing layer defines how transactions are processed, validated, and metered over the shared state. Its design dimensions include the processing model with dependency management, validation timing like correctness conditions enforced prior to execution and cost metering approach determining how computational and I/O cost are measured and charged. These choices showcase the control-flow patterns, batch strategies, and atomicity guarantees.

    \item \textbf{Ownership and capability layer:} The ownership and capability layer encodes asset-control semantics and transfer authorization. This layer specifies how ownership is represented,  what approval mechanisms exist, and whether the platform allows mandatory interception points for transfers. These choices determine how NFT ownership, permissions, royalties, and higher-level capabilities are expressed and enforced.

\end{itemize}

%Within each layer, a blockchain platform makes concrete design choices which can be referred as architectural primitives. These primitive are basic capabilities on which the NFT features are built. The application of NFT code can only use the primitive but can not create new ones.

The architectural model enables us to map each NFT feature to the specific layers and primitives on which it depends. However, the feature–primitive mapping can vary across blockchain platforms. For example, token identity may rely on a sequential numeric identifier primitive on one platform, while another platform may use a cryptographically derived identifier, resulting in different observable behaviors for the same feature. Therefore, understanding the feature–primitive mapping is essential to explain divergences in NFT behavior across platforms. This raises an additional question: Are the four layers truly independent, or do design choices at one layer constrain the options available at others?

\subsection{Inter-Layer Dependency in NFT Architecture}

%In this subsection, we show that the NFT architecture exhibits directional inter-layer dependencies and examine how these dependencies affect NFT features.

In our process of analyzing different blockchain platforms, we observed a pattern where platform that differ in design choices at lower layers tend to differ at higher layers as well. Most NFT-supporting platforms expose NFTs with broadly similar goals, yet their identity mechanisms,
ownership representations, transaction execution models, and metadata handling differ substantially.  This phenomenon is not a coincidence but reflects a structural property of layered architectures. Prior work on layered systems \cite{1265643}  argues that a layered system exhibits directional dependencies, meaning that decisions at lower layers shape, constrain, and enable the design space available to upper layers. 
Blockchain platform exhibit upward layer dependencies. A design choice or primitive at a higher layer is built using the primitives at lower layers, and therefore inherits their constraints. The platform's cryptographic choices determine how  its state is organized. The state organization then shapes what transaction models are efficient. The transaction processing model then decides the implementation of ownership patterns.  Each layer is built on capabilities and absorbs the limitations of the layers below.
%Similar stacked models have been adopted for blockchain platforms, where application-layer capabilities are analyzed in terms of underlying network, consensus and replicated-state-machine layers \cite{9239372}. We apply this principle to our NFT architecture.

The dependency structure creates a critical implication for cross-chain NFT migration. A mismatch between the primitives of different blockchain platforms at the lower layer propagates upwards.  So, if the cryptographic layer differs between the source and target platforms, the choices and features at the higher layer may be affected.  Similarly, a difference in transaction execution models (sequential vs. parallel) at the transaction-processing layer affects how features like batch operations must be re-implemented, even when both platforms support batch minting. It is, therefore, essential to understand the dependencies for analyzing the mismatch in the feature behavior.

%In our NFT architecture, the cryptographic layers established how the entities are named, what constitutes a valid address, the formation of identifiers, and the recognizable signature formats. The state-management layer decides how the data is organized, stored and access mechanism. The transaction-processing layer then determines how the transactions can be processed, and what computation can occur atomically. The higher layers must express their functionality using these established constructs. The upper layer can not have a naming convention, storage pattern or any other operational feature that lower later do not support.   

On the one hand, each layer defines foundational constructs that higher layers must use \cite{7202962}. On the other hand, the higher layers also inherit constraints along with the capabilities. For example, a platform having deterministic address derivation mechanism enables address to be computed from seeds, but it also requires that the addresses cannot be arbitrary values. This is also evident from the claims made by \cite{10.5555/3349349,TABATABAEI2023100575} that application capabilities in blockchain-based applications are bounded by foundational design choices. Thus, the inter-dependence of the layers in NFT architecture and the architectural differences at lower layers produce a significant difference in the NFT's feature behavior.
%The duality here shows that higher layer just not gain capabilities only but they have to absorb the limitations also. Previous studies on technical dependencies also confirms that foundational design choices create rules that all dependent components must follow \cite{10.1287/isre.2017.0739}.

%The higher-layer choices are composed of lower-layer primitives. A service of features at a higher layer is not a standalone abstraction but an assembly of components from multiple layers. 

\begin{figure}
    \centering
    \includegraphics[scale=0.5]{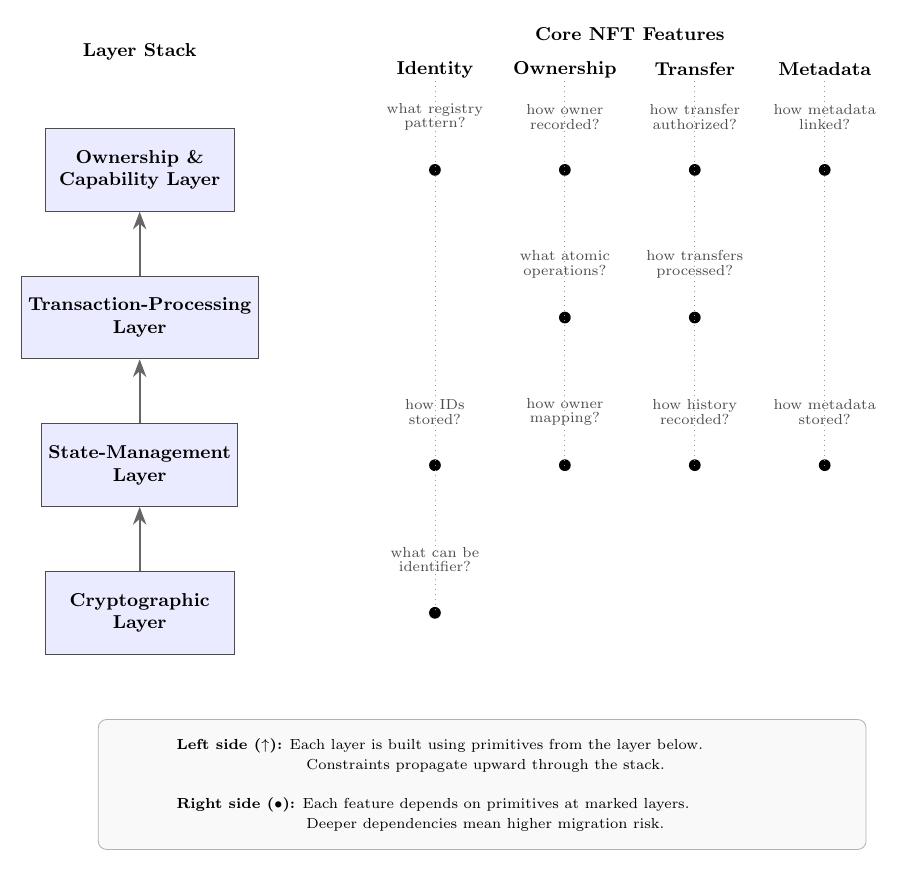}
    \caption{Inter-layer dependencies in the NFT architecture and their mapping to core NFT features}
    \label{fig:dependencu}
\end{figure}

Fig. \ref{fig:dependencu} visualizes the example relationship of inter-layer dependencies and feature implementation. The left side shows the NFT architectural stack with upward constraint propagation. The right side traces four core NFT features through the layers, marking dependencies with dots. The questions at each layer-feature intersection identify what the feature needs from that layer. Although Fig. \ref{fig:dependencu} shows only $F_{\mathit{core}}$, any additional feature $f \in F_{\mathit{ext}}$ can be analyzed using the same procedure. We need to start at the ownership and capability layer, follow its use of the transaction-processing layer, state-management layer, and cryptographic primitives along the edge $E$. At each layer, we need to record on which primitive the feature $f$ depends on. In this way, extended features such as royalties, batch operations or some other behavioral aspects of NFT can be treated as composition of same architectural primitives that support the core features and inherit upward dependencies.

To support precise reasoning in migration analysis, we formalize this layered structure. Let $L = \{\ell_{\mathit{crypto}}, \ell_{\mathit{state}}, \ell_{\mathit{tx}}, \ell_{\mathit{own}}\}$ denote the four layers. We model the architecture of a given blockchain platform as a directed acyclic graph $\mathcal{A} = (L, E)$, where an edge  $(\ell_i, \ell_j) \in E$  indicates that primitives at layer $\ell_j$ are implemented \emph{on top of} primitives at layer $\ell_i$. In our four-layer setting this yields a natural ordering
\[
\ell_{\mathit{crypto}} \rightarrow \ell_{\mathit{state}} \rightarrow \ell_{\mathit{tx}} \rightarrow \ell_{\mathit{own}}.
\]
We write $P_\ell$ for the set of primitives available at that layer on a given platform. The graph $(L, E)$ and the associated primitive sets $P_\ell$, together constitute the platform's architectural profile which are the capabilities and constraints within which each NFT feature must be implemented. 
%We use this formalization in our methodology to analyze the migration outcome.

%This architectural model and its inter-layer dependencies provide the theoretical foundation for our analysis of cross-chain NFT migration.
%In the next section, we use this model to define migration mismatches in terms of feature requirements and platform capabilities, and to derive a systematic migration analysis methodology.

\section{Cross-chain NFT Migration Analysis Methodology}\label{sec:migration-analysis}

Using the NFT architecture $\mathcal{A} = (L, E, \{P_\ell\}_{\ell \in L})$, we now formalize cross-chain NFT migration as a compatibility problem between a source NFT's feature requirements and a target platform's capabilities. We adopt a feature-centric view where a migration is \emph{successful} if all essential features of the source implementation can be reproduced on the target chain, and a \emph{migration mismatch} occurs when this is not possible due to architectural differences.

Migration feasibility then depends on whether the target platform provides the architectural primitives required to realize each feature of the source NFT. Let $C_{\mathit{src}}$ and $C_{\mathit{tgt}}$ denote the source and target blockchain platforms, with architectures
\[
\mathcal{A}_{\mathit{src}} = (L, E, \{P^{\mathit{src}}_\ell\}_{\ell \in L})
\quad\text{and}\quad
\mathcal{A}_{\mathit{tgt}} = (L, E, \{P^{\mathit{tgt}}_\ell\}_{\ell \in L}),
\]
where $L$ and $E$ are the four layers and their inter-layer dependency graph, and
$P^{\mathit{src}}_\ell$, $P^{\mathit{tgt}}_\ell$ are the sets of primitives available at layer $\ell$ on each platform.
Consider a concrete NFT implementation on $C_{\mathit{src}}$ with a set of features $F$ where $F = F_{\mathit{core}} \cup F_{\mathit{ext}}$. Note that $F_{\mathit{core}}$ must be included in the mismatch analysis, while $F_{\mathit{ext}}$ depends upon the use-case and whether the analyst wants to include them or not.

For each feature $f \in F$ we write $D_{\mathit{src}}(f) \subseteq \bigcup_{\ell \in L} P^{\mathit{src}}_\ell$
for the set of architectural primitives on the source chain that this feature actually relies on; in practice, $D_{\mathit{src}}(f)$ is obtained by tracing the feature implementation through the inter-layer structure of $\mathcal{A}_{\mathit{src}}$ (Phases~1 and~2 of our methodology described below). A cross-chain migration aims to construct a corresponding implementation of these features on $C_{\mathit{tgt}}$, using primitives drawn from $P_{\mathit{tgt}} = \bigcup_{\ell \in L} P^{\mathit{tgt}}_\ell$.

In this view, migration analysis reduces to comparing $D_{\mathit{src}}(f)$ with $P_{\mathit{tgt}}$ for each feature $f \in F$. If all required
primitives in $D_{\mathit{src}}(f)$ have suitable counterparts on the target platform, the feature can, in principle, be re-implemented with similar behavior. If some required primitives are missing or fundamentally different, the feature must be redesigned or may become impossible to reproduce. To identify potential migration issues, we introduce a systematic four-phase methodology that traces NFT
features back to their architectural dependencies on the source chain and compares them with the capabilities of candidate target chains.

\noindent\textbf{Phase 1: Source Feature Specification}. 

\noindent\textbf{Objective:}  To construct a feature specification for the source NFT implementation. Since $F_{\mathit{core}}$ must always be included,
Phase~1 (i) identifies any additional features $F_{\mathit{ext}}$ relevant for the NFT under study and (ii) records the behavioral aspects of each feature on the source chain.

\noindent\textbf{Procedure:} Initialize the feature set $F$ with $F_{\mathit{core}}$. Inspect the NFT's public interface, including contract code, emitted events, and accompanying documentation. This inspection can be done either manually or using analysis tools such as \cite{TIAN2025100303}. The inspection looks for recurring behaviors that (a) change the NFT's state or usage rules, or (b) are explicitly described as part of the NFT’s intended functionality in the documentation. Each such behavior is given a name and added to $F_{\mathit{ext}}$.
Now, for each feature  $f \in F = F_{\mathit{core}} \cup F_{\mathit{ext}}$, write a short structured specification that records, when $f$ is triggered (relevant function, inputs, or actions), how $f$ changes on-chain state or user-observable behavior, and which conditions  are intended to  hold before and after $f$ executes. 

\noindent\textbf{Output Artifact:} A source feature profile which has list of features $F = F_{\mathit{core}} \cup F_{\mathit{ext}}$ together with a  behavioral specification for each feature.

\noindent\textbf{Phase 2: Primitive Dependency Mapping.}

\noindent\textbf{Objective:} For each source feature identified in Phase 1, we derive the set of architectural primitives that the feature relies on. From Phase 1, we have the feature set, $F = F_{\mathit{core}} \cup F_{\mathit{ext}}$ and a behavioral specification for each $f \in F$. 
For every feature $f$, Phase 2 constructs a dependency set $D_{\mathit{src}}(f) \subseteq \bigcup_{\ell \in L} P^{\mathit{src}}_\ell$
containing the primitives that are required to realize $f$ as implemented on the source chain.

\noindent\textbf{Procedure:} For each feature $f \in F$, use the feature specification to find where the feature is implemented on the source chain. It can be a function, the storage location they read or write, and any other configuration parameter it uses. Next, inspect how the code and configuration interact with the four layers in $\mathcal{A}_{\mathit{src}}$.  For each such interaction, decide which primitive in $P^{\mathit{src}}_\ell$ it uses (for some  $\ell \in L$), using the layer definitions from  Section \ref{sec:nft:layerarchitecture} as a guide. Collect all identified primitives into a direct-dependency set   $D^{\mathrm{direct}}_{\mathit{src}}(f)$.
Some primitives in $D^{\mathrm{direct}}_{\mathit{src}}(f)$ are themselves built on primitives at lower layers. Using the dependency edges $E$, add to $D^{\mathrm{direct}}_{\mathit{src}}(f)$ any lower-layer primitives that these depend on. Repeat this step until no new primitives are added. 

\noindent\textbf{Output Artifact:} The result is the full dependency set $D_{\mathit{src}}(f)$ for feature $f$. Once the  $D_{\mathit{src}}(f)$ is obtained for all $f \in F$, arrange these into feature-layer matrix where each row corresponds to feature $f$ and  each column to a layer $\ell \in L$, with entries indicating which primitives in $P^{\mathit{src}}_\ell$ are required by $f$. The feature-layer matrix created here is similar to the matrix depicted in Fig. \ref{fig:dependencu}, which shows the dependency of all $ f \in F_{\mathit{core}}$ with all $l \in L$. %A complete instantiation of Phase 2 for the identity feature of an ERC-721 NFT on Ethereum is given in Section~\ref{sec:case-study}, where we show how $D_{\mathit{src}}(f)$ is constructed layer by layer.

\noindent\textbf{Phase 3: Target Platform Characterization.}

\noindent\textbf{Objective:} To characterize the target blockchain platform's architectural stack by identifying its primitive choices at each functional layer in the four-layer framework. This characterization determines which capabilities are architecturally available for implementing NFT features.
We model the target chain as $\mathcal{A}_{\mathrm{tgt}} = (L, E, \{P^{\mathrm{tgt}}_\ell\}_{\ell \in L})$ with the same layer set $L$ and dependency edges $E$  and layer-specific primitive sets $P^{\mathrm{tgt}}_\ell$ to be identified in this phase.

\noindent\textbf{Procedure:} First, identify the cryptographic primitives that govern identity establishment, integrity verification, and address derivation on the target platform. Record the  resulting set $P^{\mathrm{tgt}}_{\mathit{crypto}}$ of cryptographic  primitives. Second, identify the state-organization approach, the storage-persistence model, and the storage-pricing model. Record the set   $P^{\mathrm{tgt}}_{\mathit{state}}$ of state-management primitives.
Third, determine the transaction-processing model,  the validation timing, and the cost-metering approach. Record the set $P^{\mathrm{tgt}}_{\mathit{tx}}$ of transaction-processing primitives. Fourth, identify the supported ownership-representation model(s), transfer
  authorization mechanisms, and any mandatory transfer-interception or   hook capabilities. Record the set   $P^{\mathrm{tgt}}_{\mathit{own}}$ of ownership and capability   primitives.

\noindent\textbf{Output Artifact:} A structured target platform profile $\{P^{\mathrm{tgt}}_\ell\}_{\ell \in L}$ documenting the primitives available at each layer. We also write $P_{\mathrm{tgt}} = \bigcup_{\ell \in L} P^{\mathrm{tgt}}_\ell$ for the set of all architectural primitives on the target chain. This profile is used in Phase 4 to compare source feature requirements $D_{\mathrm{src}}(f)$ with target capabilities.

\noindent\textbf{Phase 4: Compatibility Assessment and Preservation Classification.} 

\noindent\textbf{Objective:} For each source feature, compare the primitives it requires (identified in Phase 2) against the primitives the target provides (documented in Phase 3). Based on this comparison, classify the migration outcome as \emph{natively preserved}, \emph{partial mismatch}, or \emph{complete mismatch}.
Recall that for each feature $f \in F$ we have a requirement set $D_{\mathrm{src}}(f) \subseteq \bigcup_{\ell \in L} P^{\mathrm{src}}_\ell$
from Phase~2, and that the target platform provides a primitive set $P_{\mathrm{tgt}} = \bigcup_{\ell \in L} P^{\mathrm{tgt}}_\ell$ from Phase 3. Phase 4 does not simply check whether $D_{\mathrm{src}}(f) \subseteq P_{\mathrm{tgt}}$, because the target may offer different primitives that can nonetheless ensure the same behavior of the feature. Instead, we classify the availability of each required primitive and derive the feature outcome from this classification.

\noindent\textbf{Procedure:} For each required primitive $p \in D_{\mathrm{src}}(f)$, determine how the target platform supports the corresponding functionality. We classify each $p$ into one of three cases:
\begin{itemize}[leftmargin=*]
    \item \emph{AVAILABLE}: an architecturally equivalent primitive exists in $P_{\mathrm{tgt}}$ (same role and guarantees);
    \item \emph{ALTERNATIVE}: no exact match exists, but a different primitive or combination of primitives in $P_{\mathrm{tgt}}$ can be
    used to realize the same high-level behavior, possibly with a different implementation pattern; and
    \item \emph{ABSENT}: the required functionality cannot be realized using any primitives in $P_{\mathrm{tgt}}$.
  \end{itemize}

Based on the availability classification of the primitives in   $D_{\mathrm{src}}(f)$, assign each feature $f$ to one of three outcome classes:
  \begin{description}[leftmargin=*]

    \item[\textbf{Natively preserved}:]
    All required primitives for $f$ are classified as AVAILABLE. The feature can be re-implemented on the target platform with the same primitives and, in principle, with equivalent guarantees. No feature-level mismatch is expected.

    \item[\textbf{Partial mismatch}:]
    At least one required primitive for $f$ is classified as  ALTERNATIVE, and none is ABSENT. The feature’s high-level behavior can still be preserved, but its implementation on the target platform must change (for example, using different identity formats), and some non-essential details may differ. The mismatch is at the implementation level, not at the level of overall capability.

    \item[\textbf{Complete mismatch}:]
    At least one required primitive for $f$ is classified as ABSENT.  The feature, as specified on the source chain, cannot be realized on the target platform without weakening its guarantees or changing its behavior in a fundamental way. This represents a full feature-level migration mismatch.
  \end{description}

\noindent\textbf{Output Artifact:} A preservation-outcome report that, for each feature $f \in F$, records its classification (natively preserved,
partial mismatch, or complete mismatch) together with the underlying primitive-availability reasoning. This report supports evidence-based
migration decisions by analyzing which features are expected to be preserved, which require adaptation, and which face fundamental architectural barriers before any migration is implemented.

Note that the analysis on the source chain is necessarily more detailed than on the target chain. On the source side, we have a specific NFT implementation, so we must first extract its feature set and then trace how each feature is implemented by  primitives across the layers (Phases~1–2). On the target side, by contrast, we only need a platform-wide architectural profile $\{P^{\mathrm{tgt}}_\ell\}_{\ell \in L}$. It acts as a catalog of which primitives are available at each layer. This profile can be constructed once from protocol specifications and reused for many
different NFTs. The source analysis is therefore more detailed because it captures per-NFT requirements, whereas the target analysis captures generic platform capabilities.

\section{Empirical Evaluation}\label{sec:case-study}

We evaluate our methodology using an Ethereum-to-Solana NFT migration case study, which represents a migration between two architecturally divergent platforms. We apply our four-phase methodology to a representative ERC-721 + ERC-2981 collection. This case study encompasses all four core features from $F_{\mathit{core}}$ plus three widely used extended features (royalties, batch processing, and user cryptographic identity), which collectively are sufficient to exercise all four architectural layers.

\subsection{Setup and Architectural Analysis}

\noindent\textbf{Source (Ethereum).}
We consider a standard OpenZeppelin ERC-721 implementation \cite{ERC721} with ERC-2981 royalties \cite{EIP2981}. The collection uses (i) sequential numeric token IDs, (ii) a centralized ownership mapping \texttt{mapping(uint256 $\Rightarrow$ address)}, (iii) IPFS URIs for off-chain JSON metadata, (iv) informational royalties via \texttt{royaltyInfo}, (v) loop-based batch minting, and (vi) user identities based on ECDSA over secp256k1.

\noindent\textbf{Target (Solana).}
We target Solana's SPL/Metaplex stack, characterized by (i) Ed25519 signatures and public-key or PDA-based accounts, (ii) account-based state with refundable rent, (iii) the Sealevel runtime with parallel transaction execution and compute-unit metering, and (iv) NFT ownership via distributed token accounts plus separate Metaplex metadata accounts.

Phase 1 summarizes the feature set $F$ and intended behavior of each feature on the source chain. In Phase 2, we use the NFT architectural model to map each feature $f \in F$ to the primitives it uses on Ethereum, yielding dependency sets $D_{\mathit{src}}(f) \subseteq \bigcup_{\ell \in L} P^{\mathit{src}}_\ell$. In Phase 3, we construct the corresponding architectural profile for Solana and identify, for each feature, which primitives in $D_{\mathit{src}}(f)$ have direct counterparts in $\bigcup_{\ell \in L} P^{\mathit{tgt}}_\ell$. Phase 4 then classifies each required primitive as AVAILABLE, ALTERNATIVE, or ABSENT on the target platform and derives a mismatch class.

\begin{table}[t]
\centering
\caption{Feature profile for the evaluated NFT (Phase~1 output)}
\label{tab:feature-profile}
\small
\resizebox{0.45\textwidth}{!}{
\begin{tabular}{p{3.0cm}p{6.0cm}}
\toprule
\textbf{Feature} & \textbf{Behavior on Ethereum} \\
\midrule
Identity mechanism
& Minting assigns strictly increasing numeric token IDs
  (starting from 1) and uses them as primary identifiers \\

Ownership representation
& Ownership recorded as a mapping from token ID to owner
  address; updates occur on mint, transfer, and burn \\

Transfer logic
& \texttt{transferFrom}/ \texttt{safeTransferFrom} callable
  only by owner or approved operator; effect is to update
  the owner mapping for the given token ID \\

Metadata linkage
& Per-token IPFS JSON URI stored in contract state; read
  via \texttt{tokenURI} and expected to remain stable \\

Royalty mechanism
& ERC-2981 \texttt{royaltyInfo} exposes a fixed 5\% royalty
  share; enforcement delegated to marketplaces \\

Batch operations
& Loop-based batch minting function mints multiple token IDs
  in a single transaction, subject to block gas limits \\

User cryptographic identity
& Accounts and signatures based on ECDSA over secp256k1;
  addresses derived from secp256k1 public keys \\
\bottomrule
\end{tabular}
}
\end{table}

\begin{table*}[t]
\centering
\caption{Primitive-level comparison and availability for each feature
(Ethereum $\rightarrow$ Solana, Phases~2--4 output)}
\label{tab:primitive-availability}
\resizebox{0.7\textwidth}{!}{
\small
\begin{tabular}{p{3.1cm}p{4.2cm}p{4.6cm}p{2.4cm}}
\toprule
\textbf{Feature}
& \textbf{Key source primitives $D_{\mathit{src}}(f)$}
& \textbf{Target support on Solana}
& \textbf{Mismatch class} \\
\midrule

Identity mechanism
& Sequential numeric token IDs as keys in global contract storage;
  ECDSA/secp256k1 account model
& Mint addresses derived from public keys or PDAs
  \textbf{[ALTERNATIVE to flexible numeric IDs]};
  
& Partial mismatch \\

Ownership representation
& Centralized tokenID$\rightarrow$owner mapping in a single contract;
  sequential EVM transaction processing
& Per-token SPL/ATA accounts and balances
  \textbf{[ALTERNATIVE to central registry]};
  parallel Sealevel execution
  \textbf{[ALTERNATIVE to strictly sequential processing]}
& Partial mismatch \\

Transfer logic
& ERC-721 \texttt{transferFrom}/ \texttt{safeTransferFrom} with
  owner-or-approved checks; ownership updates in the same mapping
& SPL Token \texttt{Transfer} with owner-or-delegate authority;
  ownership updates in token accounts
  \textbf{[AVAILABLE]}
& Natively preserved \\

Metadata linkage
& String URIs in contract storage; IPFS JSON metadata
& String URIs in Metaplex metadata accounts; IPFS JSON metadata
  \textbf{[AVAILABLE]}
& Natively preserved \\

Royalty mechanism
& ERC-2981 informational royalty fields; no mandatory interception
& Metaplex royalty fields (\texttt{seller\_fee\_basis\_points},
  \texttt{creators}); no mandatory interception
  \textbf{[AVAILABLE]}
& Natively preserved \\

Batch operations
& Loop-based batch mint in a single transaction; gas metering and
  block gas limit
& Multiple mint transactions over disjoint accounts; parallel execution
  and compute-unit metering
  \textbf{[ALTERNATIVE batching primitive]}
& Partial mismatch \\

User cryptographic identity
& secp256k1 private keys and ECDSA signatures; addresses from
  secp256k1 public keys
& Ed25519 keys and EdDSA signatures; no cross-curve key migration
  primitive
  \textbf{[ABSENT]}
& Complete mismatch \\

\bottomrule
\end{tabular}
}
\end{table*}

\subsection{Empirical Validation Results}
The results of Phase 1 are shown in Table~\ref{tab:feature-profile}. Results of Phases 2 to 4 are in Table~\ref{tab:primitive-availability}. To validate the identified partial and complete mismatches, we deploy functionally equivalent NFTs and observe the behavior of the studied features.

\textbf{Deployment setup.}
On Ethereum, we deploy an ERC-721 + ERC-2981 contract on a local Geth-based Sepolia setup and mint $100$ NFTs with sequential IDs and IPFS metadata URIs. On Solana Devnet, we mint $100$ corresponding NFTs using the SPL Token and Metaplex Token Metadata programs, with PDA-based mint accounts, associated token accounts for ownership, identical IPFS URIs, and 5\% royalties encoded in \texttt{seller\_fee\_basis\_points} and
\texttt{creators}. %Bridge/oracle logic is used only to coordinate migrations and is not part of the methodology; it merely illustrates the consequences of cryptographic incompatibility.

\textbf{Identity and ownership.}
As predicted, Ethereum uses numeric token IDs as primary identifiers, while Solana uses 32-byte mint addresses (PDAs or public keys). Numeric labels survive only as metadata annotations on Solana, requiring off-chain or auxiliary on-chain mappings for numeric queries. Ownership moves from a centralized mapping in contract storage to distributed token accounts, preserving token-to-owner queries in $O(1)$ time but making owner-to-tokens queries indexer-dependent. These observations match the predicted partial mismatches for identity mechanism and ownership representation.

\textbf{Batch operations.}
On Ethereum, loop-based batch minting exhibits approximately linear gas growth with batch size, bounded by block gas limits. On Solana, we realize the same logical operation through multiple transactions submitted in parallel over disjoint accounts. The Sealevel runtime executes these concurrently, yielding higher throughput. The feature is preserved at the behavioral level, but the implementation pattern and performance characteristics change, matching the predicted partial mismatch for batch operations.

\textbf{User cryptographic identity.}
The cryptographic mismatch between secp256k1 and Ed25519 prevents any reuse of Ethereum private keys on Solana. Users must generate fresh Ed25519 key pairs; cross-chain ownership correspondence is established via a trusted oracle that links Ethereum signatures to Solana addresses. This confirms the predicted complete mismatch at the cryptographic layer and the resulting reliance on trust-based coordination.

Metadata linkage and royalty mechanism behave as predicted with native preservation: IPFS URIs are carried over to Metaplex URI fields unchanged, and ERC-2981 parameters map to Metaplex royalty fields with identical economic semantics in this setting. 
All observed outcomes match the predictions.  This case study, therefore, provides evidence that architecture-centric reasoning is sufficient to anticipate when NFT features will be natively preserved, partially mismatched, or completely mismatched during cross-chain migration.

% In this Ethereum-to-Solana setting, only user cryptographic identity exhibits a complete mismatch; designing and evaluating additional complete-mismatch scenarios (for example, protocol-enforced features that depend on absent primitives) is an interesting direction for future work.

\section{Discussions}
\label{sec:discussion}

Unlike Wormhole \cite{11185063} and LayerZero’s ONFT  \cite{lawrence2025bridging}, our feature-centric methodology enables reasoning about feature-level semantics within an ICS-721 or ONFT deployment determining which semantics remain equivalent and where deeper architectural differences necessitate transformations. That said, this study,  validates the methodology on a single source–target pair  and a limited set of features. We plan to evaluate the methodology with other families of chains, such as IBC/ICS-721 transfers, object-centric platforms like Sui and Aptos, richer Ethereum standards like ERC-1155, ERC-4907 rentals. Our feature–primitive mappings are currently curated by hand. However, we view them as a precursor to an automated analysis tool that would extract dependencies from smart contracts, match them against machine-readable platform profiles, and report compatibility classes. Our current analysis focuses on on-chain architectural primitives. Off-chain infrastructure and governance  can further affect the behavior of certain features, so extension to accomdate socio-economic layers will be the natural next step. Prior to this work, there was no structured way to reason about NFT migration compatibility, as different cases demonstrate. Our framework provides this missing reasoning structure.
% , and can be extended to accommodate socio-economic layers as well.

\section{Conclusion and Future Work}
\label{sec:conclusion}

This paper reframes cross-chain NFT migration as an architecture-level compatibility problem, which is beyond moving tokens or messages between chains. The core challenge is to decide whether the target chain can realize the same NFT features that existed on the source chain. We propose a four-phase NFT cross-chain migration analysis methodology, supported by our novel proposal of a four-layer NFT architecture. The proposed methodology offers NFT creators, marketplaces, and protocol designers a structured tool for assessing which features will survive migration, which require redesign, and where trust-based mechanisms (such as oracles) are unavoidable. We evaluated the methodology using a representative Ethereum–to–Solana NFT migration case study and identified partial and complete mismatches. Our future work will be to extend the evaluation and automate the methodology.

\bibliographystyle{ieeetr}
\bibliography{ref} 

\end{document}